%% file: TADA.tex
\pdfoutput=1
\documentclass[sigconf]{acmart}

\copyrightyear{2027}
\acmYear{2027}
\setcopyright{cc}
\setcctype{by-nc-nd}
\acmConference[ASPDAC '27]{32nd Asia and South Pacific Design Automation Conference}{January 25--28, 2027}{Tokyo, Japan}
\acmBooktitle{32nd Asia and South Pacific Design Automation Conference (ASPDAC '27), January 25--28, 2027, Tokyo, Japan}
\acmDOI{10.1145/3840404.3848163}
\acmISBN{979-8-4007-2923-2/2027/01}

\usepackage{multirow}

\begin{document}

\title{TADA: Technology Aware DNN Accelerator Performance and Energy Estimator}

\settopmatter{authorsperrow=4}
\author{Tanvi Sharma}
\authornote{Tanvi Sharma is currently with NVIDIA and Surya Selvam is with Waymo. This work was carried out during their PhD program at Purdue University.}
\affiliation{%
  \institution{Purdue University}
  \city{West Lafayette}
  \country{USA}}
\email{sharm418@purdue.edu}
\orcid{0000-0001-8307-5785}

\author{Surya Selvam}
\authornotemark[1]
\affiliation{%
  \institution{Purdue University}
  \city{West Lafayette}
  \country{USA}}
\email{selvams@purdue.edu}
\orcid{0000-0002-0300-5478}

\author{Anand Raghunathan}
\affiliation{%
  \institution{Purdue University}
  \city{West Lafayette}
  \country{USA}}
\email{araghu@purdue.edu}
\orcid{0000-0002-4624-564X}

\author{Kaushik Roy}
\affiliation{%
  \institution{Purdue University}
  \city{West Lafayette}
  \country{USA}}
\email{kaushik@purdue.edu}
\orcid{0000-0002-0735-9695}

\begin{abstract}
\input{content/00-abstract}
\end{abstract}

\ccsdesc[500]{Hardware~Emerging technologies}
\ccsdesc[500]{Computer systems organization~Neural networks}
\ccsdesc[300]{Hardware~Semiconductor memory}
\ccsdesc[300]{Computing methodologies~Machine learning}

\keywords{memory technology, hardware design space exploration, machine learning
  model, data-driven approach, transformers}

\maketitle

\setlength{\textfloatsep}{8pt plus 2pt minus 2pt}
\setlength{\intextsep}{8pt plus 2pt minus 2pt}
\setlength{\floatsep}{6pt plus 2pt minus 2pt}
\setlength{\belowcaptionskip}{2pt}

\input{content/01-introduction}

\input{content/02-background}
\input{content/03-approach}
\input{content/04-evaluation}
\input{content/05-conclusion}
\input{content/06-acknowledgments}

\bibliographystyle{ACM-Reference-Format}
\bibliography{refs}

\end{document}

%% file: content/00-abstract.tex
Deep neural network (DNN) accelerators for edge inference must meet tight energy, latency, and area (EDP) constraints, and their optimal hardware architecture and dataflow configuration is typically identified through design space exploration (DSE). However, on-chip memory technology including cell type, bank organization, and process node, is rarely treated as a first-class citizen in DSE. For example, increasing memory capacity can reduce memory accesses through greater data reuse, but at the cost of higher area and access energy, thereby limiting the compute units budget. Consequently, memory technology can fundamentally shift the optimal accelerator architecture for a given area and EDP budget.

Incorporating memory technology into existing Machine Learning (ML)-based DSE methods requires repeating simulation, data generation, and model training from scratch for each technology. We present \textit{TADA}, a memory \textit{t}echnology-\textit{a}ware \textit{DNN} \textit{a}ccelerator performance and energy estimator that predicts system latency and energy for a given workload, hardware configuration, and memory technology. TADA is trained on a synthetic technology dataset designed to generalize across memory technologies. The dataset captures nearly exhaustive current and diverse future technology trends while enforcing physics-based guardrails to preserve realistic area--energy trade-offs. During training, each memory technology is represented by a \textit{technology signature}, a vector of read energy, write energy, and storage capacity characterized across a range of areas, which is provided as input alongside the workload and hardware configuration. Across 21 unseen technology curves spanning SRAM, STTRAM, ReRAM, PCM, FeFET, eDRAM, and CTT, \textit{TADA} predicts system energy and latency with 1.5--4.4\% and 0.7--1.2\% mean absolute percentage error, respectively. In three pairwise case studies, it matches the simulator-selected memory technology at 98.7--100\% of area budgets, with a mean EDP loss of at most 0.21\%, while achieving over $10^{4}\times$ higher evaluation throughput.

%% file: content/01-introduction.tex
\section{Introduction}

\begin{figure}[!htb]
    \centering
    \includegraphics[width=\linewidth]{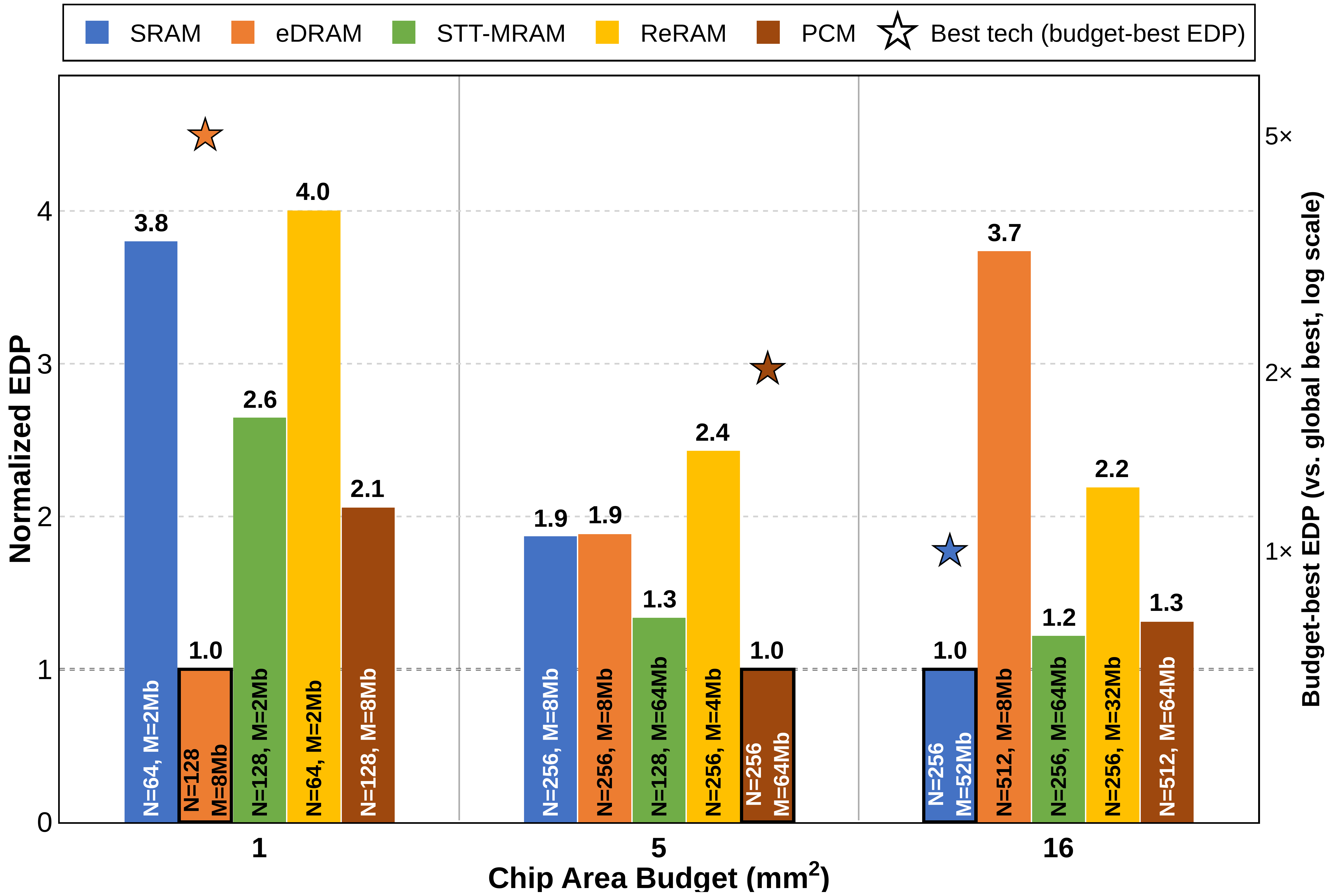}
    \caption{Best EDP for each technology after jointly optimizing compute array size and buffer capacity at each area budget for a transformer-based prefill workload. Bars are normalized to the best technology at each budget; the dashed line shows that the budget-best EDP decreases as area increases.}
    \Description{Grouped bar chart of normalized energy--delay product for each memory technology at several area budgets, with a dashed line showing the best achievable EDP falling as the area budget grows. The lowest-EDP technology changes with budget: eDRAM at 1 mm squared, PCM at 5 mm squared, and SRAM at 16 mm squared.}
    \vspace{-4mm}
    \label{fig:intro}
\end{figure}

Deep neural networks (DNNs), particularly transformers, are increasingly deployed on edge platforms with limited resources for tasks ranging from perception in real time to language modeling.
Unlike datacenter inference, edge inference must run under tight energy, latency, and area budgets, which has driven the design of specialized DNN accelerators.
Designing such an accelerator is broadly a joint design problem across three axes: the compute architecture (the type and size of the PE array), the scratchpad memory (its capacity, bandwidth, and underlying memory technology), and the dataflow mapping of a workload onto the hardware (tile size and order of spatial and temporal execution).
The compute architecture and the mapping have been studied extensively in the design space exploration (DSE) literature, but the choice of memory technology for the scratchpad buffers has received comparatively little attention, despite its strong influence on total energy, performance, and area.

The memory-macro implementation, including its bit cell, bank organization, and process node, determines buffer access energy, latency, and capacity per unit area. Dense memories such as eDRAM, STTRAM, and ReRAM reduce area but generally incur higher access energy or latency\cite{nguyen2023mcaimem, chakraborty2020resistive,pentecost2021nvmexplorer}. Within SRAM, a read-decoupled 8T cell trades larger cell area for improved read stability, thereby supporting low-voltage design compared to conventional differential-read 6T SRAM~\cite{chang2005stable}. Moreover, the read-write asymmetry and other trade-offs are usually implementation-dependent, making memory technology a key design variable for optimizing EDP under an area budget.

A memory technology, however, is not a single design point or a categorical choice with a fixed ranking.
Each technology provides an area-dependent set of memory macros whose capacity, access time and per-access read and write energies vary together.
Fig.~\ref{fig:intro} demonstrates that selecting a macro also changes the area available for the PE array, and hence, technology, buffer capacity, and compute-array size must be explored jointly. As a result, eDRAM has the lowest EDP at 1\,mm\textsuperscript{2}, PCM at 5\,mm\textsuperscript{2}, and SRAM at 16\,mm\textsuperscript{2} area budgets, respectively.
Strict area budget picks high density memory despite lower energy efficiency to maintain sufficient reuse and performance with maximum possible compute array size. At relaxed area budgets, large compute array and memory capacity that holds the working set can fit within area budget for multiple technologies, reducing the EDP difference between SRAM, STT-MRAM and PCM.
Thus, no memory technology is universally best for DNN accelerator design - its benefit depends jointly on the area budget, the required EDP limit, hardware architecture and the workload configuration.

DSE frameworks based on machine learning accelerate the DNN accelerator search space containing $\mathcal{O}(10^{12}\text{--}10^{17})$ configurations~\cite{kao2022digamma, seo2025airchitect} by learning performance and energy from hardware and mapping configurations~\cite{krishnan2023archgym, samajdar2023airchitect, seo2025airchitect, kumar2021data, huang2022learning}.
Yet, state of the art DSE efforts fix the memory technology used to generate their training data, so supporting a new technology requires new simulation and retraining.
A transferable estimator must therefore solve two problems: representing the continuous area--energy--capacity behavior of a technology rather than one operating point, and covering the technology space without access to every proprietary memory model or simulating every technology of interest.

To tackle the above challenges, we present \textbf{TADA}, a \emph{technology-aware DNN accelerator} performance and energy estimator.
For representation, we describe each technology by a \emph{technology signature}, a vector of its read energy, write energy, and storage capacity characterized across a range of areas, and provide it to the predictor as input alongside the workload and hardware configuration. Therefore, our predictor model operates over a continuous technology space and can be generalized to unseen technologies.
For covering the proprietary technology space, we generate a synthetic dataset by sampling around a small set of characterized reference technologies and progressively sampling beyond the boundary of the real technology range. We apply physical guardrails to keep every sample plausible, and filtering to keep the set compact while maintaining diversity.
We then use architectural simulator that includes dataflow mapping search to generate training dataset using this synthetic corpus. A regression model is trained on the simulator generated dataset to predict total system latency and energy. Although the methodology applies to any accelerator based on compute arrays with a scratchpad memory hierarchy, we instantiate and evaluate it on digital compute-in-memory (DCIM) designs that show potential for adoption in highly energy-efficient edge DNN inference.

The key contributions of this work are as follows:
\begin{enumerate}
    \item \textbf{Technology-aware DNN accelerator estimator:} We propose TADA, a generalizable ML estimator that rapidly evaluates DNN accelerator configurations across heterogeneous memory technologies under varying area budgets, running over $10^{4}\times$ faster than a parallelized architectural analytical simulator.
    \item \textbf{Synthetic technology dataset:} We develop a data generation and curation pipeline that produces physically plausible technology signatures capturing realistic tradeoffs between area density and access energy, enabling a single model to generalize to unseen technologies without simulating each technology separately and without access to proprietary device data.
    \item \textbf{Experimental validation:} We evaluate TADA on 21 unseen technology curves spanning SRAM, STTRAM, ReRAM, PCM, FeFET, eDRAM, and CTT, achieving 1.5--4.4\% energy and 0.7--1.2\% latency MAPE ($R^2\geq0.998$). Across three pairwise case studies, TADA matches the simulator's technology or bank-organization choice at 98.7--100\% of area budgets with mean EDP loss of at most 0.21\%.
\end{enumerate}

%% file: content/02-background.tex
\section{Background and Related Works} \label{sec:background}

\begin{figure*}[t]
    \centering
    \includegraphics[width=\textwidth,trim={0 16pt 0 0},clip]{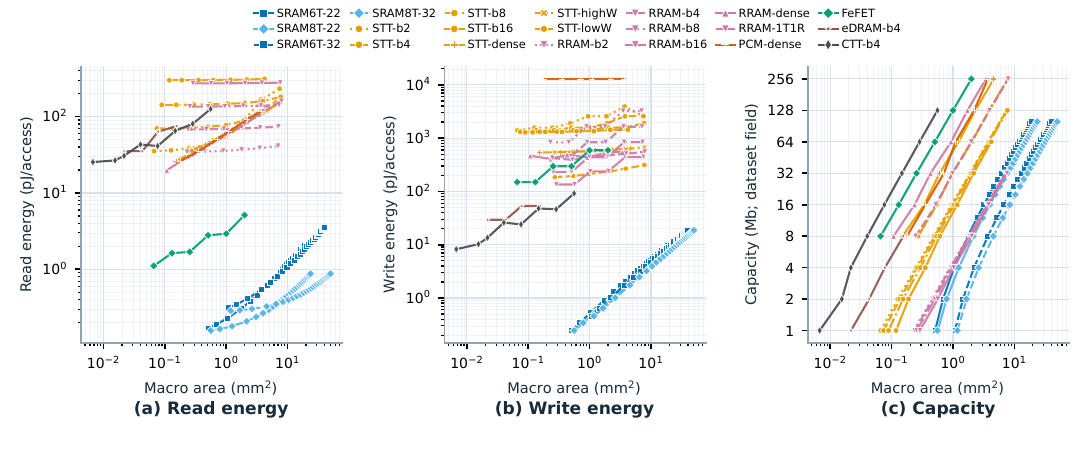}
    \captionsetup{skip=2pt}
    \caption{Area-dependent signatures for 21 evaluation curves, combining CACTI, NVSim/NVMExplorer, and independently characterized inputs~\cite{muralimanohar2009cacti,dong2012nvsim,pentecost2021nvmexplorer}.}
    \Description{Three log-scale plots show read energy, write energy, and capacity versus modeled macro area for 21 evaluation curves.}
    \label{fig:real-tech-curves}
\end{figure*}

Modern DNNs increasingly rely on transformer architectures, whose blocks combine attention with a feed-forward network (MLP). Linear projections first form the Query (Q), Key (K), and Value (V) matrices. Attention then computes the attention scores through $QK^T$, applies Softmax to obtain normalized weights, and multiplies these weights by V. The projection, attention, and MLP operations are dominated by batched general matrix multiplications (GEMMs), whereas Softmax, GELU, and Layer Normalization primarily involve vector operations and reductions. In autoregressive inference, \textit{prefill} processes multiple input tokens simultaneously, exposing substantial matrix--matrix parallelism and typically operating in a compute-bound regime. In contrast, token-by-token \textit{decode} repeatedly accesses model weights and the key--value (KV) cache, making it often limited by memory bandwidth and data movement~\cite{wu2026memexplorer}.

These workloads are commonly executed on spatial accelerators, where arrays of processing elements (PEs) for matrix and vector operations communicate through a global buffer or on-chip interconnect. Tensor Processing Units (TPUs), for example, implement matrix multiplication using systolic arrays~\cite{jouppi2017tpu}, while Eyeriss employs an array of processing elements with a row-stationary dataflow~\cite{chen2016eyeriss}. Later accelerators introduced more flexible distribution and reduction networks to support multiple dataflows and improve adaptability across workloads~\cite{kwon2021heterogeneous, kwon2018maeri}.

Compute-in-memory (CIM) architectures instead perform multiply--accumulate operations within or near the memory array, reducing the movement of data between storage and compute units. Analog and digital CIM occupy vary in terms of numerical accuracy, energy efficiency, and dataflow flexibility. Digital CIM (DCIM) uses digital arithmetic and reduction structures to preserve exact integer computation~\cite{houshmand2023benchmarking, sharma2025compute}. A DCIM macro integrates digital multiply--accumulate and reduction logic within an SRAM array, thereby reducing data movement while retaining deterministic integer arithmetic. In this work, we use a DCIM macro as the MatMul unit; however, the proposed methodology generalizes to other PE-array-based accelerators, where PE denotes a processing element.

\subsection{Memory Technology and Design Space Exploration}

DNN accelerators use a memory hierarchy in which an on-chip global buffer captures reuse and reduces expensive off-chip DRAM traffic~\cite{chen2016eyeriss, kwon2021heterogeneous}.
The selected memory technology determines buffer capacity, silicon area, access time and the energy of each read and write. \textit{Technology here could refer to memory cell type, array organization, inter-bank network, process node or any optimization that modifies the memory macro characteristics.} Fig.~\ref{fig:real-tech-curves} illustrates 21 different memory technology curves such as SRAM, STT-MRAM, ReRAM, PCM, and FeFET with 2048-bit width. As area increases, capacity and per-access energy change together. Note that the non-volatile memory options are not uniformly better or worse, because their density and read--write asymmetry depend on the cell and array organization~\cite{pentecost2021nvmexplorer}.
Because DNNs reuse buffered data heavily, memory choice strongly shapes system-level energy and area under a fixed budget. As an example, the system-level choices for a sample transformer-based prefill workload are shown in Fig.~\ref{fig:intro} across area budgets. 
This co-dependence and widespread characteristics of memory-macro motivates for an ultra-fast estimator that predicts accelerator performance and energy across technologies without re-simulating and retraining for each candidate.

Designing an efficient DNN accelerator requires a joint search over compute, memory, and mapping choices.
ML-based frameworks increasingly automate this search.
One such framework, PRIME, learns a conservative offline objective for accelerator optimization~\cite{kumar2021data}, while VAESA learns a continuous, reconstructible latent hardware space~\cite{huang2022learning}.
ArchGym couples search algorithms such as Bayesian optimization and ML algorithms to architecture simulators for DSE~\cite{krishnan2023archgym}.
AIRCHITECT predicts an optimal accelerator architecture configuration and mapping as a classification problem and v2 trains an encoder-decoder transformer with unified ordinal representations~\cite{samajdar2023airchitect,seo2025airchitect}.
NAPEL learns high-level performance and energy models for near-memory architectures and evaluates transfer to unseen applications~\cite{singh2019napel}.
Together, these methods search or predict across workloads, constraints, mappings, and hardware configurations, but do not consider an unseen buffer technology with an area--energy--capacity curve. 

Note, however, memory-aware accelerator exploration itself is not new.
The closest precedent for proposed TADA framework is the work by Li \emph{et al.}, who co-explore SRAM, eDRAM, MRAM, and vertical RRAM as on-chip storage for mobile DNN accelerators~\cite{li2019onchip}.
For analog CIM, He \emph{et al.} jointly explore macro/analog-digital-converter configurations and SRAM, PCM, ReRAM, ferroelectric, and spintronic memory families under iso-area budgets~\cite{he2022memory}.
NVMExplorer compares characterized embedded non-volatile memories across device, array, system, and application levels~\cite{pentecost2021nvmexplorer}, while MemExplorer parameterizes heterogeneous memory and NPU choices for prefill and decode~\cite{wu2026memexplorer}. However, TADA estimator can rapidly evaluate a new or existing memory technology for transformer-based workloads without any new technology-specific system simulations or re-training. 

%% file: content/03-approach.tex
\vspace{-2.5mm}
\section{Our Approach: TADA} \label{sec:main}

\begin{figure*}[t]
    \centering
    \includegraphics[width=\linewidth]{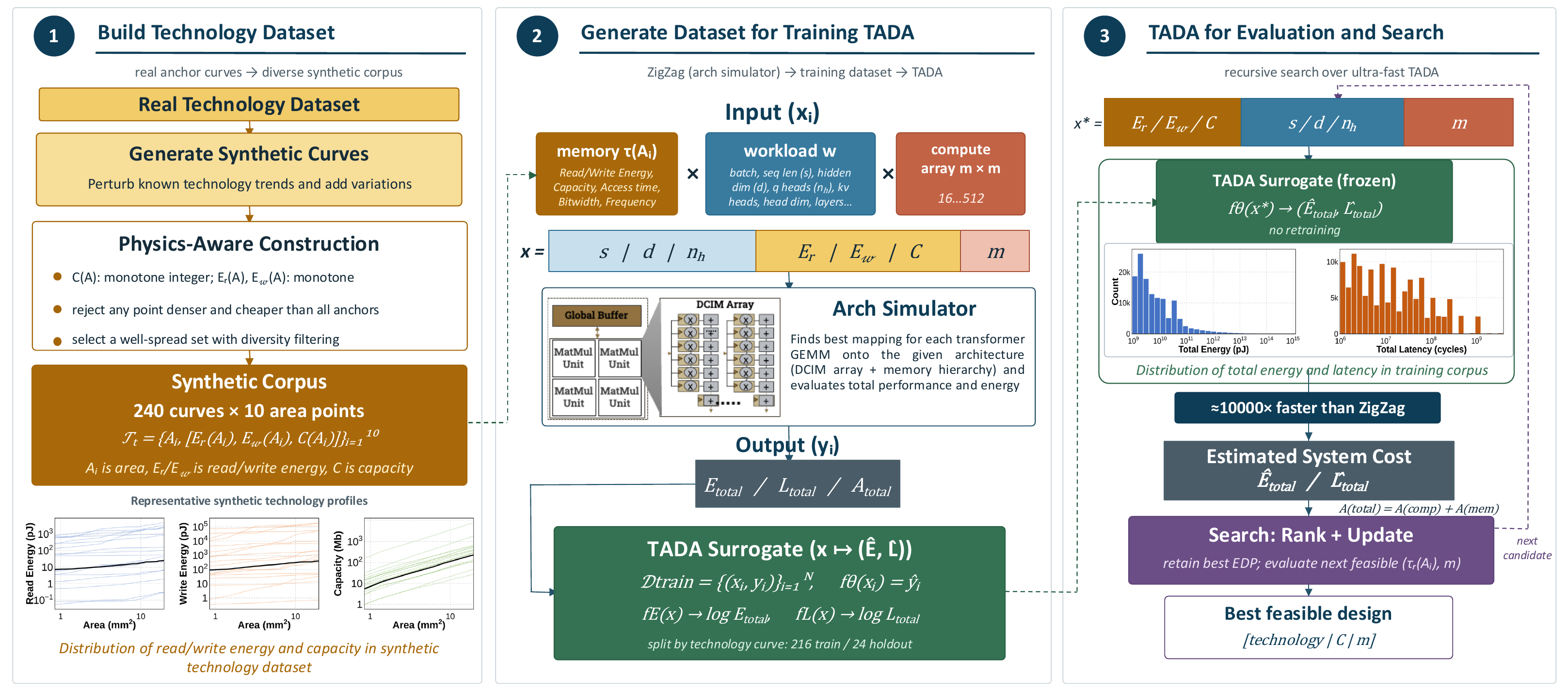}
    \caption{Overview of TADA. \textbf{(1)} Synthetic technology dataset is created by using real technology for baseline sampling, then perturbed and filtered by physics-aware constraints to form a diverse synthetic corpus; the inset shows representative read-energy, write-energy, and capacity profiles. \textbf{(2)} Each technology signature is combined with a workload and DCIM-array configuration, labeled by ZigZag, and used to train the TADA surrogate. \textbf{(3)} At evaluation, the frozen surrogate estimates energy and latency for unseen technology--workload--hardware candidates and supports iterative area-constrained search; the inset summarizes the energy and latency distribution of the simulator-labeled training corpus.}
    \Description{Three-stage flow diagram. Stage one generates synthetic memory technology curves from real reference technologies, perturbs them, and filters them with physics-aware constraints. Stage two combines each technology signature with workload and DCIM array configurations, labels them with the ZigZag simulator, and trains a regression surrogate. Stage three uses the trained surrogate to estimate energy and latency for unseen technologies and to search for the best configuration under an area budget.}
    \label{fig:tada_framework}
\end{figure*}

TADA is a technology-aware DNN accelerator performance and energy predictor.
It addresses the two challenges identified in the introduction: representing memory technology as a continuous design variable and obtaining sufficient technology diversity to predict for technologies unseen during training.
TADA represents a memory technology by its macro-level characteristics alongside the workload and accelerator configurations, and learns a shared predictor from a technology-diverse training dataset generated using the synthetic corpus.
Fig.~\ref{fig:tada_framework} gives an overview of our approach, including synthetic technology data generation, training a regression model using the dataset generated from an architectural simulator, and using TADA estimator for hardware configuration search.

\vspace{-1mm}
\subsection{Problem Formulation}

For a workload descriptor $w$, a non-memory accelerator configuration $h$, and a memory-technology signature $\tau$, TADA learns a function
\begin{equation}
f_\theta(w,h,\tau)=(\hat{E},\hat{L}),
\end{equation}
where $\hat{E}$ and $\hat{L}$ are the predicted total system energy and latency, respectively.
Unlike conventional accelerator predictors, $\tau$ is not fixed: the same function must model how different memory characteristics change the system-level outcome for a given workload and accelerator configuration.
Because $\tau$ represents physical macro characteristics rather than a categorical technology identifier, the learned function can be applied to technology profiles that are absent from the labeled training set.

\vspace{-3mm}
\subsection{Synthetic Technology Space}

Each reference technology is represented by three curves that describe read energy, write energy, and storage capacity as functions of memory area.
We generate complete technology profiles rather than independent operating points because changing the macro area jointly changes its capacity and access-energy characteristics.
Candidate profiles are constructed by varying both the magnitude and area-scaling behavior observed in the reference curves, thereby expanding the technology space while retaining realistic area-dependent trends.

We stratify this space according to the strength of the available evidence.
The characterized region covers trends observed directly in reference technologies, the documented region extends coverage to trends supported by published device classes, and the extrapolated region provides limited coverage beyond those ranges.
Sampling becomes progressively sparser as the supporting evidence weakens.
We retain profiles only when capacity and access energy follow physically plausible area scaling and the profile does not unrealistically dominate all reference technologies.
Diversity selection then removes near-duplicate profiles so that the training set spans distinct area--energy--capacity trade-offs.

\subsection{Technology Signature}

For a memory technology $t$, we define its technology profile as
\begin{equation}
\mathcal{P}_t =
\left\{
\left(
A_{t,i},
E_{r,t}(A_{t,i}),
E_{w,t}(A_{t,i}),
C_t(A_{t,i})
\right)
\right\}_{i=1}^{K_t},
\end{equation}
where $A_{t,i}$ is the area of the $i$th candidate macro, $E_{r,t}$ and $E_{w,t}$ are its read and write energies, and $C_t$ is its storage capacity.
Each operating point in this profile produces the signature
\begin{equation}
\tau_{t,i} =
\left(
E_{r,t}(A_{t,i}),
E_{w,t}(A_{t,i}),
C_t(A_{t,i})
\right).
\end{equation}
The macro area $A_{t,i}$ determines whether the candidate satisfies the accelerator area budget, while $\tau_{t,i}$ is provided to the predictor with the workload and accelerator configuration.

This representation captures how access energy and storage capacity change together as memory area increases.
Unlike a categorical technology label, it allows the same predictor to compare macros from different technologies and evaluate a newly characterized profile without technology-specific retraining.

\subsection{Technology-Aware Performance Estimation and Search}

For each workload, accelerator configuration, and candidate memory macro, a reference system-level evaluation provides the corresponding total energy and latency.
These evaluations form the labeled dataset
\begin{equation}
\mathcal{D} =
\left\{
\left(w_j,h_j,\tau_j,E_j,L_j\right)
\right\}_{j=1}^{N}.
\end{equation}
TADA trains one shared predictor across all technology profiles by solving
\begin{equation}
\theta^* =
\arg\min_{\theta}
\sum_{j=1}^{N}
\mathcal{L}
\left(
f_\theta(w_j,h_j,\tau_j),
(E_j,L_j)
\right),
\end{equation}
where $\mathcal{L}$ denotes the regression loss.
Sharing the predictor across technologies enables it to learn how memory-access characteristics interact with the workload and accelerator configuration instead of learning a separate model for each technology.

For a total-area budget $B$, the feasible design set is
\begin{equation}
\mathcal{F}_B =
\left\{
(h,t,i)
\;\middle|\;
A_h(h)+A_{t,i}\leq B
\right\},
\end{equation}
where $A_h(h)$ is the area of the non-memory accelerator components.
TADA ranks the feasible candidates using their predicted energy--delay product and selects
\begin{equation}
(h^*,t^*,i^*) =
\arg\min_{(h,t,i)\in\mathcal{F}_B}
\hat{E}(w,h,\tau_{t,i})
\hat{L}(w,h,\tau_{t,i}).
\end{equation}
Thus, a single trained estimator jointly supports memory-technology and accelerator-configuration selection under different area budgets.

%% file: content/04-evaluation.tex
\section{Evaluation} \label{sec:results}

\subsection{Experimental Setup}

We evaluate TADA on nine prefill transformer GEMM workloads with embedding dimensions of 768 and 1024, 12 and 16 attention heads, and sequence lengths from 128 to 2048.
The hardware space comprises six DCIM array sizes from $16\times16$ to $512\times512$ and ten buffer-capacity operating points per technology curve, with total accelerator area up to approximately 20\,mm\textsuperscript{2}.

The training corpus contains 240 synthetic technology curves and 129{,}600 workload--hardware--technology configurations.
We label each configuration with the ZigZag analytical simulator~\cite{houshmand2023benchmarking, mei2021zigzag}; the resulting sweep comprises 1{,}555{,}200 simulations with no failures.
We split the corpus by technology curve, assigning 90\% of curves to training and reserving the remaining 10\% for synthetic holdout evaluation.
This split prevents configurations of a held-out technology from appearing in training.
The labeled configurations span several orders of magnitude in energy and latency, as summarized in the third panel of Fig.~\ref{fig:tada_framework}.
On the synthetic holdout, TADA achieves 3.8\% energy MAPE and 3.1\% latency MAPE.

We evaluate transfer on 21 real or independently characterized technology curves that are absent from training.
The test set includes CACTI-characterized SRAM6T and SRAM8T at 22 and 32\,nm, and NVSim-characterized STTRAM and ReRAM at $2\times2$, $4\times4$, and $8\times8$ bank organizations~\cite{muralimanohar2009cacti,dong2012nvsim}.
For SRAM8T, we use a 0.3\,V supply assumption enabled by the decoupled read path, yielding approximately $0.375\times$ the SRAM6T read energy at 22\,nm with 50\% sparsity, at the cost of a larger cell (185\,F\textsuperscript{2} versus 140\,F\textsuperscript{2}).
The remaining classes are PCM, FeFET, eDRAM, additional STTRAM and ReRAM cell variants, and a charge-trap-transistor (CTT) memory; the PCM, FeFET, and variant designs were characterized independently of our flow with cell parameters that differ from every reference curve.
All test configurations have independent simulator labels, totaling 123{,}660 simulations with no failures.
\vspace{-2mm}

\begin{table}[t]
\centering
\caption{TADA accuracy on unseen memory technologies. MAPE is computed in linear space and $R^2$ in log space, consistent with the model target.}
\label{tab:results}
\resizebox{\linewidth}{!}{
\begin{tabular}{lcccc}
\hline
\multirow{2}{*}{\textbf{Technology}} & \multicolumn{2}{c}{\textbf{Total Energy}} & \multicolumn{2}{c}{\textbf{Total Latency}} \\
 & MAPE & $R^2$ & MAPE & $R^2$ \\
\hline
SRAM 6T/8T (22\,nm)          & 1.5\% & 0.9996 & 0.7\% & 0.9999 \\
SRAM 6T/8T (32\,nm)          & 1.5\% & 0.9996 & 0.7\% & 0.9999 \\
STTRAM (7 designs)           & 2.8\% & 0.9988 & 1.0\% & 0.9999 \\
ReRAM (6 designs)            & 2.8\% & 0.9989 & 1.0\% & 0.9999 \\
PCM                          & 4.4\% & 0.9980 & 0.9\% & 0.9999 \\
FeFET                        & 1.8\% & 0.9995 & 0.9\% & 0.9999 \\
eDRAM                        & 3.2\% & 0.9985 & 1.2\% & 0.9998 \\
CTT                          & 2.5\% & 0.9991 & 1.0\% & 0.9999 \\
\hline
\end{tabular}
}
\end{table}

\subsection{Accuracy on Unseen Technologies}

Table~\ref{tab:results} shows that TADA predicts total energy within 1.5--4.4\% MAPE and total latency within 0.7--1.2\% MAPE for every in-scope technology class.
The corresponding $R^2$ values are at least 0.9980 for energy and 0.9998 for latency.
CTT is excluded entirely from the reference set and exhibits read--write asymmetry unlike most reference technologies, yet TADA attains 2.5\% energy MAPE and 1.0\% latency MAPE.
PCM and FeFET, characterized independently of our flow, are predicted with 4.4\% and 1.8\% energy MAPE, respectively.

\subsection{Memory Technology DSE using TADA}

\begin{figure}[!t]
    \centering
    \vspace{-4mm}
    \includegraphics[width=\linewidth]{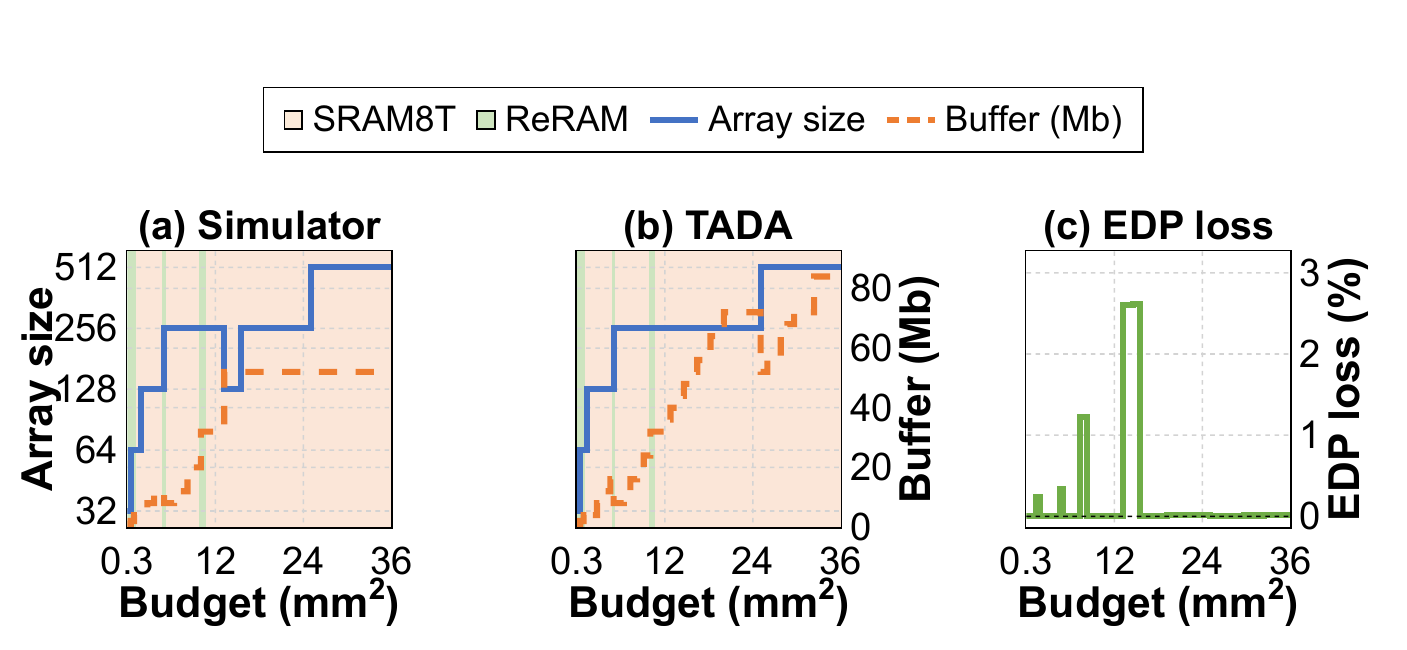}
    \caption{Best-EDP configuration versus area budget for SRAM8T versus ReRAM: (a) simulator, (b) TADA, and (c) EDP loss of TADA's selection.}
    \Description{Three panels plotted against area budget. Panels (a) and (b) show the compute array size and buffer capacity selected by the simulator and by TADA, with background shading marking whether SRAM8T or ReRAM is chosen; the selected technology alternates five times and TADA reproduces every alternation. Panel (c) shows TADA's EDP loss, averaging 0.21 percent.}
    \label{fig:sramrramhw}
\end{figure}

\begin{figure}[!t]
    \centering
    \vspace{-4mm}
    \includegraphics[width=\linewidth]{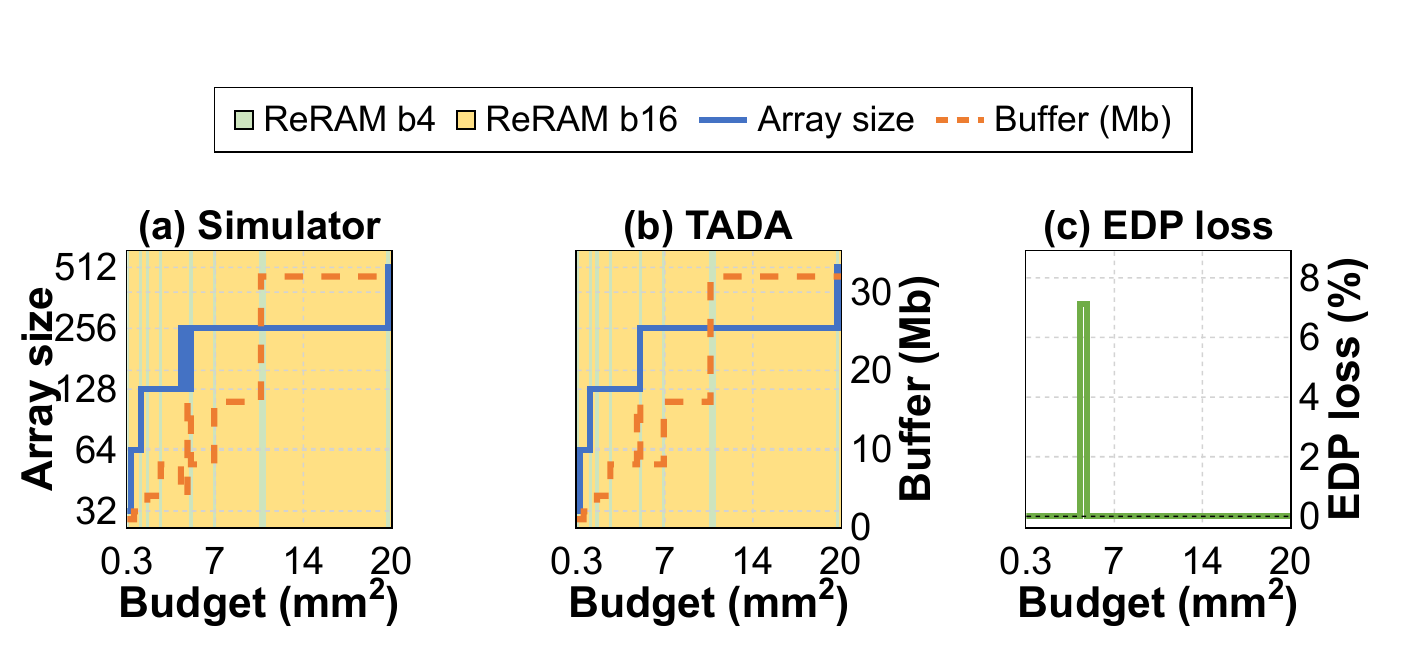}
    \caption{Best-EDP configuration versus area budget for ReRAM bank-4 versus bank-16 organizations: (a) simulator, (b) TADA, and (c) EDP loss of TADA's selection.}
    \Description{Three panels plotted against area budget. Panels (a) and (b) show the compute array size and buffer capacity selected by the simulator and by TADA, with background shading marking whether the ReRAM bank-4 or bank-16 organization is chosen; the choice alternates fifteen times and TADA matches it at every budget. Panel (c) shows TADA's EDP loss, averaging 0.18 percent with one spike near 7 percent at a transition budget.}
    \vspace{-4mm}
    \label{fig:bankinghw}
\end{figure}

Although TADA is a predictor, a useful downstream test is whether its estimates preserve EDP-based technology and hardware selections.
For each workload and area budget, we compare the configuration selected from TADA's predictions with the simulator-optimal configuration in Fig.~\ref{fig:sttrramhw}-\ref{fig:bankinghw}. The graphs show how the optimal array size and buffer capacity changes as area budget increases along with the choice of memory technology depicted by the background color.
We also report the EDP gap between the simulator-evaluated system energy and latency relative to the hardware configuration that TADA selected.
Exact agreement is intentionally stringent when several configurations have nearly identical EDP, making TADA estimation nearly optimal but ultra-fast.
We report three pairwise case studies---STTRAM versus ReRAM, SRAM8T versus ReRAM, and ReRAM bank-4 versus bank-16.

\subsubsection{STTRAM versus ReRAM}

\begin{figure}[!t]
    \centering
    \vspace{-4mm}
    \includegraphics[width=\linewidth]{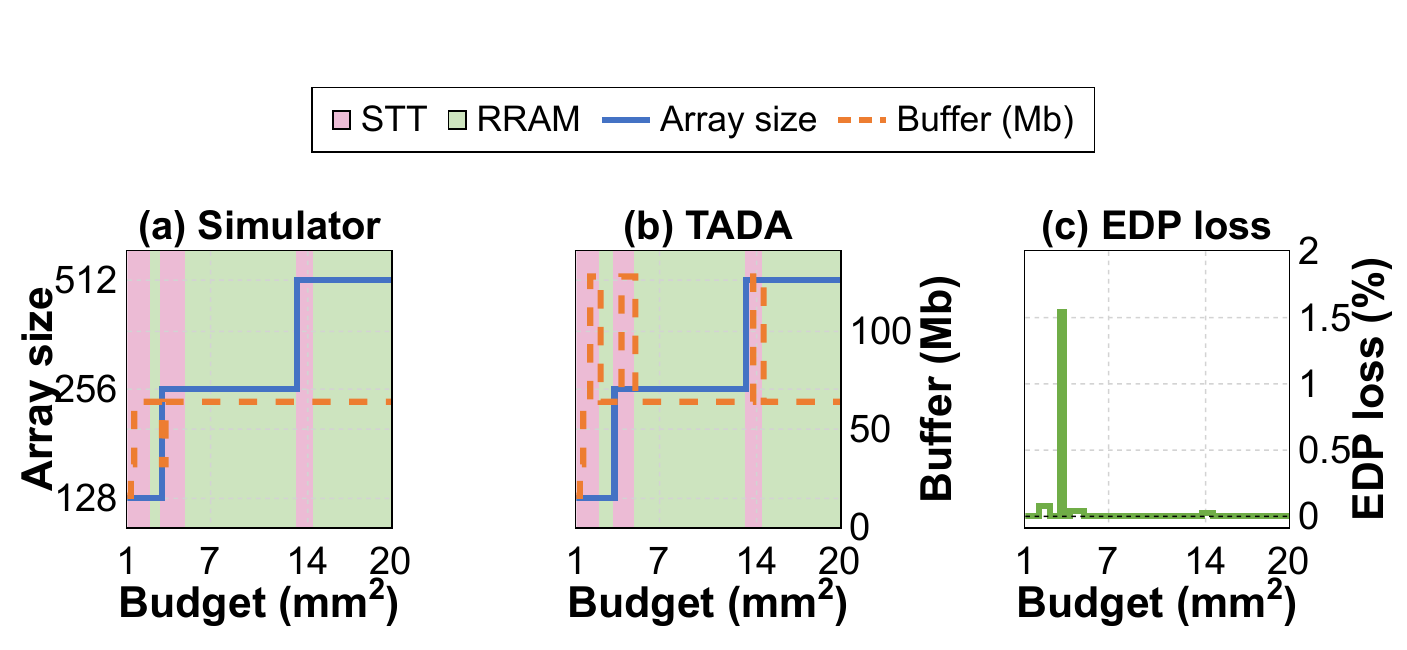}
    \caption{Best-EDP configuration versus area budget for STTRAM/ReRAM: (a) simulator, (b) TADA, and (c) EDP loss of TADA's selection.}
    \Description{Three panels plotted against area budget. Panels (a) and (b) show the compute array size and buffer capacity selected by the simulator and by TADA, with background shading marking whether STTRAM or ReRAM is chosen; the two panels closely match. Panel (c) shows TADA's EDP loss, which stays near zero with a maximum of about 1.5 percent.}
    \label{fig:sttrramhw}
\end{figure}

For a transformer with multi-head attention and sequence length 512, hidden dimension 768, and 12 heads, TADA matches the selected technology at 98.7\% of budgets and the full array--buffer configuration at 85.7\% of budgets (Fig.~\ref{fig:sttrramhw}). The mean EDP loss is 0.03\%, and the maximum is 1.5\%. Fig.~\ref{fig:real-tech-curves} shows STT occupies lower area than RRAM but consumes higher read and write energy. Owing to its high area density, STT is preferred at lower area budget with $128\times128$ array size and a buffer capacity of 20Mb. As area budget expands, first the buffer capacity increases to improve data reuse and hence reduce EDP. Then, memory technology shifts to RRAM because it can support similar hardware configuration with lower read/write energy. Again, as area budget expands to $5mm^2$, STT becomes the preferred technology because it can provide sufficient memory capacity while expanding the array size to $256\times256$. The other case studies also follow similar reasoning.

\subsubsection{SRAM8T versus ReRAM}

Choosing between a volatile SRAM buffer and a denser non-volatile alternative is a canonical technology decision.
For SRAM8T versus ReRAM (Fig.~\ref{fig:sramrramhw}), the simulator-optimal technology alternates five times across the budget sweep as each technology reaches its next array or capacity tier first.
TADA reproduces every alternation, matching the selected technology at 100\% of budgets with a mean EDP loss of 0.21\%.

\subsubsection{ReRAM Banking-Organization Selection}

Technology selection also extends \emph{within} a technology class: the bank organization changes read energy, write energy, and area density, and the optimal organization is budget-dependent.
For ReRAM at bank-4 versus bank-16 (Fig.~\ref{fig:bankinghw}), the simulator-optimal choice alternates fifteen times across the sweep.
TADA matches the selected organization at every budget and the exact array--buffer configuration at 97.5\% of budgets; the mean EDP loss is 0.18\%, with a single transition budget reaching 7\%.

\subsection{Runtime and Predictor Analysis}

Table~\ref{tab:runtime} reports independent same-host measurements on an Apple M4 and on an AMD EPYC 9334 server container; TADA and ZigZag are restricted to one CPU thread in both environments.
The measured speedups of TADA over ZigZag are $5.51\times10^{4}$ on the M4 and $4.50\times10^{4}$ on the EPYC as shown in Table~\ref{tab:runtime}, the values being arithmetic means over 100 equivalent repetitions.

\begin{table}[t]
\centering
\caption{Direct same-host runtime over 148 matched evaluations; entries are arithmetic means.}
\label{tab:runtime}
\small
\begin{tabular}{@{}lcc@{}}
\toprule
\textbf{Quantity} & \textbf{Apple M4} & \textbf{EPYC 9334} \\
\midrule
ZigZag time/config. (s) & 23.21 & 38.90 \\
TADA time/config. (ms) & 0.421 & 0.864 \\
Direct speedup & $5.51\times10^{4}$ & $4.50\times10^{4}$ \\
\bottomrule
\end{tabular}
\end{table}

\begin{figure}
    \centering
    \includegraphics[width=\linewidth]{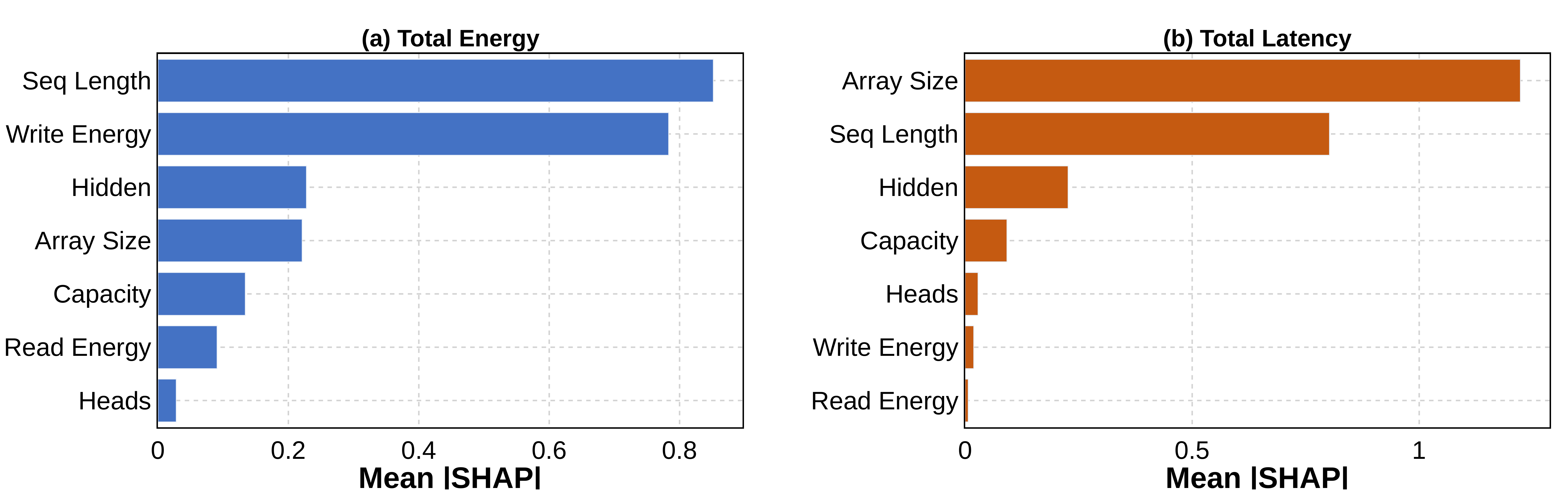}
    \caption{Mean absolute SHAP attributions for total energy (a) and total latency (b).}
    \Description{Two SHAP feature-importance plots. For total energy (a), sequence length and write energy are the most influential features, followed by hidden dimension, array size, and capacity. For total latency (b), array size and workload shape dominate, with capacity contributing through data reuse.}
    \vspace{-4mm}
    \label{fig:shap}
\end{figure}

SHAP (SHapley Additive exPlanations) attributions in Fig.~\ref{fig:shap} show that sequence length and write energy are the leading features for total-energy prediction, followed by hidden dimension, array size, and capacity. Latency is dominated by array size and workload shape, with capacity contributing through data reuse. The graphs show an inverse relationship of energy and latency with array size, indicating that larger arrays generally lead to reduced energy/cycles per operation. Additionally, when capacity falls below a certain threshold, total energy increases as capacity decreases because of low data reuse. Capacity has a stronger impact on latency than on energy because larger capacities reduce data re-loading, improving execution efficiency.

%% file: content/05-conclusion.tex
\section{Conclusion} \label{sec:conclusion}

This work presents TADA, a technology-aware performance and energy estimator that makes on-chip memory technology an explicit, continuous variable in DNN accelerator design space exploration. 
By representing each technology as a signature of its read-write energy and area density characteristics and training on an almost exhaustive synthetic dataset, a single model predicts system-level energy and latency within 1.5--4.4\% and 0.7--1.2\% MAPE across 21 unseen technology curves spanning SRAM, STTRAM, ReRAM, PCM, FeFET, eDRAM, and CTT.
Across three pairwise case studies, TADA matches the simulator's technology or bank-organization choice at 98.7--100\% of area budgets with mean EDP loss of at most 0.21\%, while providing over $10^{4}\times$ higher configuration-evaluation throughput. The work also yields two reusable artifacts: a synthetic technology dataset for future ML-based DSE methods and a lightweight estimator for rapidly screening the system-level impact of emerging memory technologies. Future work will focus on expanding technology signature, architecture configurations and using workload-level EDP targets to derive the memory-macro characteristics.

%% file: content/06-acknowledgments.tex
\begin{acks}
This work was supported in part by the Center for the Co-Design of Cognitive Systems (CoCoSys), one of the seven centers in JUMP 2.0, a Semiconductor Research Corporation (SRC) program sponsored by DARPA.
The authors would also like to sincerely acknowledge Prof.~Sumeet Kumar Gupta for his generous help and insightful discussions.
\end{acks}

%% file: TADA.bbl

\begin{thebibliography}{23}


\ifx \showCODEN    \undefined \def \showCODEN     #1{\unskip}     \fi
\ifx \showDOI      \undefined \def \showDOI       #1{#1}\fi
\ifx \showISBNx    \undefined \def \showISBNx     #1{\unskip}     \fi
\ifx \showISBNxiii \undefined \def \showISBNxiii  #1{\unskip}     \fi
\ifx \showISSN     \undefined \def \showISSN      #1{\unskip}     \fi
\ifx \showLCCN     \undefined \def \showLCCN      #1{\unskip}     \fi
\ifx \shownote     \undefined \def \shownote      #1{#1}          \fi
\ifx \showarticletitle \undefined \def \showarticletitle #1{#1}   \fi
\ifx \showURL      \undefined \def \showURL       {\relax}        \fi
\providecommand\bibfield[2]{#2}
\providecommand\bibinfo[2]{#2}
\providecommand\natexlab[1]{#1}
\providecommand\showeprint[2][]{arXiv:#2}

\bibitem[Chakraborty et~al\mbox{.}(2020)]%
        {chakraborty2020resistive}
\bibfield{author}{\bibinfo{person}{Indranil Chakraborty}, \bibinfo{person}{Mustafa Ali}, \bibinfo{person}{Aayush Ankit}, \bibinfo{person}{Shubham Jain}, \bibinfo{person}{Sourjya Roy}, \bibinfo{person}{Shrihari Sridharan}, \bibinfo{person}{Amogh Agrawal}, \bibinfo{person}{Anand Raghunathan}, {and} \bibinfo{person}{Kaushik Roy}.} \bibinfo{year}{2020}\natexlab{}.
\newblock \showarticletitle{Resistive crossbars as approximate hardware building blocks for machine learning: Opportunities and challenges}.
\newblock \bibinfo{journal}{\emph{Proc. IEEE}} \bibinfo{volume}{108}, \bibinfo{number}{12} (\bibinfo{year}{2020}), \bibinfo{pages}{2276--2310}.
\newblock


\bibitem[Chang et~al\mbox{.}(2005)]%
        {chang2005stable}
\bibfield{author}{\bibinfo{person}{Leland Chang}, \bibinfo{person}{David~M Fried}, \bibinfo{person}{Jack Hergenrother}, \bibinfo{person}{Jeffrey~W Sleight}, \bibinfo{person}{Robert~H Dennard}, \bibinfo{person}{Robert~K Montoye}, \bibinfo{person}{Lidija Sekaric}, \bibinfo{person}{Sharee~J McNab}, \bibinfo{person}{Anna~W Topol}, \bibinfo{person}{Charlotte~D Adams}, {et~al\mbox{.}}} \bibinfo{year}{2005}\natexlab{}.
\newblock \showarticletitle{Stable SRAM cell design for the 32 nm node and beyond}. In \bibinfo{booktitle}{\emph{Digest of Technical Papers. 2005 Symposium on VLSI Technology, 2005.}} IEEE, \bibinfo{pages}{128--129}.
\newblock


\bibitem[Chen et~al\mbox{.}(2016)]%
        {chen2016eyeriss}
\bibfield{author}{\bibinfo{person}{Yu-Hsin Chen}, \bibinfo{person}{Joel Emer}, {and} \bibinfo{person}{Vivienne Sze}.} \bibinfo{year}{2016}\natexlab{}.
\newblock \showarticletitle{Eyeriss: A spatial architecture for energy-efficient dataflow for convolutional neural networks}.
\newblock \bibinfo{journal}{\emph{ACM SIGARCH computer architecture news}} \bibinfo{volume}{44}, \bibinfo{number}{3} (\bibinfo{year}{2016}), \bibinfo{pages}{367--379}.
\newblock


\bibitem[Dong et~al\mbox{.}(2012)]%
        {dong2012nvsim}
\bibfield{author}{\bibinfo{person}{Xiangyu Dong}, \bibinfo{person}{Cong Xu}, \bibinfo{person}{Yuan Xie}, {and} \bibinfo{person}{Norman~P Jouppi}.} \bibinfo{year}{2012}\natexlab{}.
\newblock \showarticletitle{NVSim: A circuit-level performance, energy, and area model for emerging nonvolatile memory}.
\newblock \bibinfo{journal}{\emph{IEEE Transactions on Computer-Aided Design of Integrated Circuits and Systems}} \bibinfo{volume}{31}, \bibinfo{number}{7} (\bibinfo{year}{2012}), \bibinfo{pages}{994--1007}.
\newblock


\bibitem[He et~al\mbox{.}(2022)]%
        {he2022memory}
\bibfield{author}{\bibinfo{person}{Kang He}, \bibinfo{person}{Indranil Chakraborty}, \bibinfo{person}{Cheng Wang}, {and} \bibinfo{person}{Kaushik Roy}.} \bibinfo{year}{2022}\natexlab{}.
\newblock \showarticletitle{Design Space and Memory Technology Co-Exploration for In-Memory Computing Based Machine Learning Accelerators}. In \bibinfo{booktitle}{\emph{Proceedings of the 41st IEEE/ACM International Conference on Computer-Aided Design}}. \bibinfo{publisher}{ACM}, \bibinfo{pages}{1--9}.
\newblock
\urldef\tempurl%
\url{https://doi.org/10.1145/3508352.3549453}
\showDOI{\tempurl}


\bibitem[Houshmand et~al\mbox{.}(2023)]%
        {houshmand2023benchmarking}
\bibfield{author}{\bibinfo{person}{Pouya Houshmand}, \bibinfo{person}{Jiacong Sun}, {and} \bibinfo{person}{Marian Verhelst}.} \bibinfo{year}{2023}\natexlab{}.
\newblock \showarticletitle{Benchmarking and modeling of analog and digital {SRAM} in-memory computing architectures}.
\newblock \bibinfo{journal}{\emph{arXiv preprint arXiv:2305.18335}} (\bibinfo{year}{2023}).
\newblock


\bibitem[Huang et~al\mbox{.}(2022)]%
        {huang2022learning}
\bibfield{author}{\bibinfo{person}{Qijing Huang}, \bibinfo{person}{Charles Hong}, \bibinfo{person}{John Wawrzynek}, \bibinfo{person}{Mahesh Subedar}, {and} \bibinfo{person}{Yakun~Sophia Shao}.} \bibinfo{year}{2022}\natexlab{}.
\newblock \showarticletitle{Learning a continuous and reconstructible latent space for hardware accelerator design}. In \bibinfo{booktitle}{\emph{2022 IEEE International Symposium on Performance Analysis of Systems and Software (ISPASS)}}. IEEE, \bibinfo{pages}{277--287}.
\newblock


\bibitem[Jouppi et~al\mbox{.}(2017)]%
        {jouppi2017tpu}
\bibfield{author}{\bibinfo{person}{Norman~P. Jouppi}, \bibinfo{person}{Cliff Young}, \bibinfo{person}{Nishant Patil}, \bibinfo{person}{David Patterson}, {et~al\mbox{.}}} \bibinfo{year}{2017}\natexlab{}.
\newblock \showarticletitle{In-Datacenter Performance Analysis of a {Tensor Processing Unit}}. In \bibinfo{booktitle}{\emph{2017 ACM/IEEE 44th Annual International Symposium on Computer Architecture (ISCA)}}. \bibinfo{pages}{1--12}.
\newblock
\urldef\tempurl%
\url{https://doi.org/10.1145/3079856.3080246}
\showDOI{\tempurl}


\bibitem[Kao et~al\mbox{.}(2022)]%
        {kao2022digamma}
\bibfield{author}{\bibinfo{person}{Sheng-Chun Kao}, \bibinfo{person}{Michael Pellauer}, \bibinfo{person}{Angshuman Parashar}, {and} \bibinfo{person}{Tushar Krishna}.} \bibinfo{year}{2022}\natexlab{}.
\newblock \showarticletitle{DiGamma: Domain-aware genetic algorithm for HW-mapping co-optimization for DNN accelerators}. In \bibinfo{booktitle}{\emph{2022 Design, Automation \& Test in Europe Conference \& Exhibition (DATE)}}. IEEE, \bibinfo{pages}{232--237}.
\newblock


\bibitem[Krishnan et~al\mbox{.}(2023)]%
        {krishnan2023archgym}
\bibfield{author}{\bibinfo{person}{Srivatsan Krishnan}, \bibinfo{person}{Amir Yazdanbakhsh}, \bibinfo{person}{Shvetank Prakash}, \bibinfo{person}{Jason Jabbour}, \bibinfo{person}{Ikechukwu Uchendu}, \bibinfo{person}{Susobhan Ghosh}, \bibinfo{person}{Behzad Boroujerdian}, \bibinfo{person}{Daniel Richins}, \bibinfo{person}{Devashree Tripathy}, \bibinfo{person}{Aleksandra Faust}, {et~al\mbox{.}}} \bibinfo{year}{2023}\natexlab{}.
\newblock \showarticletitle{{ArchGym}: An open-source gymnasium for machine learning assisted architecture design}. In \bibinfo{booktitle}{\emph{Proceedings of the 50th Annual International Symposium on Computer Architecture}}. \bibinfo{pages}{1--16}.
\newblock


\bibitem[Kumar et~al\mbox{.}(2022)]%
        {kumar2021data}
\bibfield{author}{\bibinfo{person}{Aviral Kumar}, \bibinfo{person}{Amir Yazdanbakhsh}, \bibinfo{person}{Milad Hashemi}, \bibinfo{person}{Kevin Swersky}, {and} \bibinfo{person}{Sergey Levine}.} \bibinfo{year}{2022}\natexlab{}.
\newblock \showarticletitle{Data-driven offline optimization for architecting hardware accelerators}. In \bibinfo{booktitle}{\emph{International Conference on Learning Representations}}.
\newblock


\bibitem[Kwon et~al\mbox{.}(2021)]%
        {kwon2021heterogeneous}
\bibfield{author}{\bibinfo{person}{Hyoukjun Kwon}, \bibinfo{person}{Liangzhen Lai}, \bibinfo{person}{Michael Pellauer}, \bibinfo{person}{Tushar Krishna}, \bibinfo{person}{Yu-Hsin Chen}, {and} \bibinfo{person}{Vikas Chandra}.} \bibinfo{year}{2021}\natexlab{}.
\newblock \showarticletitle{Heterogeneous dataflow accelerators for multi-{DNN} workloads}. In \bibinfo{booktitle}{\emph{2021 IEEE International Symposium on High-Performance Computer Architecture (HPCA)}}. IEEE, \bibinfo{pages}{71--83}.
\newblock


\bibitem[Kwon et~al\mbox{.}(2018)]%
        {kwon2018maeri}
\bibfield{author}{\bibinfo{person}{Hyoukjun Kwon}, \bibinfo{person}{Ananda Samajdar}, {and} \bibinfo{person}{Tushar Krishna}.} \bibinfo{year}{2018}\natexlab{}.
\newblock \showarticletitle{{MAERI}: Enabling flexible dataflow mapping over {DNN} accelerators via reconfigurable interconnects}.
\newblock \bibinfo{journal}{\emph{ACM Sigplan Notices}} \bibinfo{volume}{53}, \bibinfo{number}{2} (\bibinfo{year}{2018}), \bibinfo{pages}{461--475}.
\newblock


\bibitem[Li et~al\mbox{.}(2019)]%
        {li2019onchip}
\bibfield{author}{\bibinfo{person}{Haitong Li}, \bibinfo{person}{Mudit Bhargava}, \bibinfo{person}{Paul~N. Whatmough}, {and} \bibinfo{person}{H.-S.~Philip Wong}.} \bibinfo{year}{2019}\natexlab{}.
\newblock \showarticletitle{On-Chip Memory Technology Design Space Explorations for Mobile Deep Neural Network Accelerators}. In \bibinfo{booktitle}{\emph{Proceedings of the 56th Annual Design Automation Conference 2019}}. \bibinfo{publisher}{ACM}, \bibinfo{pages}{1--6}.
\newblock
\urldef\tempurl%
\url{https://doi.org/10.1145/3316781.3317874}
\showDOI{\tempurl}


\bibitem[Mei et~al\mbox{.}(2021)]%
        {mei2021zigzag}
\bibfield{author}{\bibinfo{person}{Linyan Mei}, \bibinfo{person}{Pouya Houshmand}, \bibinfo{person}{Vikram Jain}, \bibinfo{person}{Sebastian Giraldo}, {and} \bibinfo{person}{Marian Verhelst}.} \bibinfo{year}{2021}\natexlab{}.
\newblock \showarticletitle{{ZigZag}: Enlarging Joint Architecture-Mapping Design Space Exploration for {DNN} Accelerators}.
\newblock \bibinfo{journal}{\emph{IEEE Trans. Comput.}} \bibinfo{volume}{70}, \bibinfo{number}{8} (\bibinfo{year}{2021}), \bibinfo{pages}{1160--1174}.
\newblock
\urldef\tempurl%
\url{https://doi.org/10.1109/TC.2021.3059962}
\showDOI{\tempurl}


\bibitem[Muralimanohar et~al\mbox{.}(2009)]%
        {muralimanohar2009cacti}
\bibfield{author}{\bibinfo{person}{Naveen Muralimanohar}, \bibinfo{person}{Rajeev Balasubramonian}, {and} \bibinfo{person}{Norman~P Jouppi}.} \bibinfo{year}{2009}\natexlab{}.
\newblock \bibinfo{booktitle}{\emph{CACTI 6.0: A tool to model large caches}}.
\newblock \bibinfo{type}{{T}echnical {R}eport} HPL-2009-85. \bibinfo{institution}{HP Laboratories}.
\newblock


\bibitem[Nguyen et~al\mbox{.}(2023)]%
        {nguyen2023mcaimem}
\bibfield{author}{\bibinfo{person}{Duy-Thanh Nguyen}, \bibinfo{person}{Abhiroop Bhattacharjee}, \bibinfo{person}{Abhishek Moitra}, {and} \bibinfo{person}{Priyadarshini Panda}.} \bibinfo{year}{2023}\natexlab{}.
\newblock \showarticletitle{MCAIMem: a Mixed SRAM and eDRAM Cell for Area and Energy-efficient on-chip AI Memory}.
\newblock \bibinfo{journal}{\emph{arXiv preprint arXiv:2312.03559}} (\bibinfo{year}{2023}).
\newblock


\bibitem[Pentecost et~al\mbox{.}(2022)]%
        {pentecost2021nvmexplorer}
\bibfield{author}{\bibinfo{person}{Lillian Pentecost}, \bibinfo{person}{Alexander Hankin}, \bibinfo{person}{Marco Donato}, \bibinfo{person}{Mark Hempstead}, \bibinfo{person}{Gu-Yeon Wei}, {and} \bibinfo{person}{David Brooks}.} \bibinfo{year}{2022}\natexlab{}.
\newblock \showarticletitle{{NVMExplorer}: A framework for cross-stack comparisons of embedded non-volatile memories}. In \bibinfo{booktitle}{\emph{2022 IEEE International Symposium on High-Performance Computer Architecture (HPCA)}}. \bibinfo{pages}{938--956}.
\newblock
\urldef\tempurl%
\url{https://doi.org/10.1109/HPCA53966.2022.00073}
\showDOI{\tempurl}


\bibitem[Samajdar et~al\mbox{.}(2023)]%
        {samajdar2023airchitect}
\bibfield{author}{\bibinfo{person}{Ananda Samajdar}, \bibinfo{person}{Jan~Moritz Joseph}, {and} \bibinfo{person}{Tushar Krishna}.} \bibinfo{year}{2023}\natexlab{}.
\newblock \showarticletitle{{AIRCHITECT}: Automating Hardware Architecture and Mapping Optimization}. In \bibinfo{booktitle}{\emph{2023 Design, Automation \& Test in Europe Conference \& Exhibition (DATE)}}. IEEE, \bibinfo{pages}{1--6}.
\newblock


\bibitem[Seo et~al\mbox{.}(2025)]%
        {seo2025airchitect}
\bibfield{author}{\bibinfo{person}{Jamin Seo}, \bibinfo{person}{Akshat Ramachandran}, \bibinfo{person}{Yu-Chuan Chuang}, \bibinfo{person}{Anirudh Itagi}, {and} \bibinfo{person}{Tushar Krishna}.} \bibinfo{year}{2025}\natexlab{}.
\newblock \showarticletitle{{AIRCHITECT} v2: Learning the Hardware Accelerator Design Space Through Unified Representations}. In \bibinfo{booktitle}{\emph{2025 Design, Automation \& Test in Europe Conference (DATE)}}. IEEE, \bibinfo{pages}{1--7}.
\newblock


\bibitem[Sharma et~al\mbox{.}(2025)]%
        {sharma2025compute}
\bibfield{author}{\bibinfo{person}{Tanvi Sharma}, \bibinfo{person}{Mustafa Ali}, \bibinfo{person}{Indranil Chakraborty}, {and} \bibinfo{person}{Kaushik Roy}.} \bibinfo{year}{2025}\natexlab{}.
\newblock \showarticletitle{What, When, Where to Compute-in-Memory for Efficient Matrix Multiplication during Machine Learning Inference}.
\newblock \bibinfo{journal}{\emph{IEEE Transactions on Emerging Topics in Computing}} \bibinfo{volume}{13}, \bibinfo{number}{3} (\bibinfo{year}{2025}), \bibinfo{pages}{1215--1229}.
\newblock
\urldef\tempurl%
\url{https://doi.org/10.1109/TETC.2025.3574508}
\showDOI{\tempurl}


\bibitem[Singh et~al\mbox{.}(2019)]%
        {singh2019napel}
\bibfield{author}{\bibinfo{person}{Gagandeep Singh}, \bibinfo{person}{Juan G\'{o}mez-Luna}, \bibinfo{person}{Giovanni Mariani}, \bibinfo{person}{Geraldo~F. Oliveira}, \bibinfo{person}{Stefano Corda}, \bibinfo{person}{Sander Stuijk}, \bibinfo{person}{Onur Mutlu}, {and} \bibinfo{person}{Henk Corporaal}.} \bibinfo{year}{2019}\natexlab{}.
\newblock \showarticletitle{{NAPEL}: Near-Memory Computing Application Performance Prediction via Ensemble Learning}. In \bibinfo{booktitle}{\emph{Proceedings of the 56th Annual Design Automation Conference 2019}}. \bibinfo{publisher}{ACM}, \bibinfo{pages}{1--6}.
\newblock
\urldef\tempurl%
\url{https://doi.org/10.1145/3316781.3317867}
\showDOI{\tempurl}


\bibitem[Wu et~al\mbox{.}(2026)]%
        {wu2026memexplorer}
\bibfield{author}{\bibinfo{person}{Haoran Wu}, \bibinfo{person}{Zeyu Cao}, \bibinfo{person}{Yao Lai}, \bibinfo{person}{Binglei Lou}, \bibinfo{person}{Jiayi Nie}, \bibinfo{person}{Can Xiao}, \bibinfo{person}{Timi Adeniran}, \bibinfo{person}{Przemyslaw Forys}, \bibinfo{person}{Kauser Johar}, \bibinfo{person}{Catriona Wright}, \bibinfo{person}{Junyi Liu}, \bibinfo{person}{Kai Shi}, \bibinfo{person}{Nicholas~D. Lane}, \bibinfo{person}{Rika Antonova}, \bibinfo{person}{Jianyi Cheng}, \bibinfo{person}{Timothy Jones}, \bibinfo{person}{Aaron Zhao}, {and} \bibinfo{person}{Robert Mullins}.} \bibinfo{year}{2026}\natexlab{}.
\newblock \showarticletitle{{MemExplorer}: Navigating the Heterogeneous Memory Design Space for Agentic Inference {NPUs}}.
\newblock \bibinfo{journal}{\emph{arXiv preprint arXiv:2604.16007}} (\bibinfo{year}{2026}).
\newblock


\end{thebibliography}
